# On the Dynamical Ferromagnetic, Quantum Hall, and Relativistic Effects on the Carbon Nanotubes Nucleation and Growth Mechanism

**Reginald B. Little***
National High Magnetic Field Laboratory
Florida State University
Tallahassee, Florida
*Corresponding Author: Tel: 850-644-0311, Email: redge_little@yahoo.com

**Alexandru S. Biris**
University of Arkansas at Little Rock, Applied Science Department,
Nanotechnology Center, 2801 S. University Ave, Little Rock, Arkansas
Tel: 501-749-9148, Email: asbiris@ualr.edu

**Abstract**:

The mechanism of carbon nanotube (CNT) nucleation and growth has been a mystery for 15 years, since the discovery of this most explored material of the $20^{th}$ century. Prior models have attempted the extension of classical transport mechanisms used to explain the older, bigger, micron-sized filamentous carbon formations. In July 2000, a more thorough, detailed, and accurate nonclassical, relativistic mechanism was formulated considering the detailed dynamics of the electronics of spin and rehybridization between the carbon and catalyst via novel mesoscopic phenomena driven by intrinsic dynamical ferromagnetic fields by spin currents and spin waves of the catalyst for activating catalytically stimulated, synchronized, orchestrated, simultaneous, and coherent hydrocarbon decomposition, adsorption, absorption, transport, electronic subshell rehybridization by spin mechanics, and multi-atomic bond rearrangements for the nucleation and growth of the CNT. In this dynamical magnetic mechanism, quantum mechanical effects and relativistic effects of intense many body spin-orbital interactions for novel orbital hybrid dynamics (Little Effect) were proposed due to the mesoscopic size of the system. The formulation of this dynamical ferromagnetic mechanism naturally led to the first realization and explanation of a physical basis for ferromagnetic nano-carbon. This discovered ferromagnetism of the carbon and forming CNT intermediates facilitates the coupling of the CNT to the ferrocatalyst for the spin currents and spin waves of the catalyst to organize and synchronize the 12 steps for the nucleation and growth processes of the CNT. Here this dynamic ferromagnetic mechanism of CNT formation via spin currents and spin waves is proven by imposing both an external static magnetic field via the bore of a strong DC magnet and an external dynamic magnetic field via intense radio frequency electromagnetic radiation for influencing the proposed spin currents and spin waves for the observations of the static magnetic field's hindrance on the CNT formation and the dynamic magnetic field's enhancement and selectivity of CNT nucleation and growth.

**Introduction**:

The single wall carbon nanotubes (SWCNT) were discovered in 1993 by Sumio Iijima [1] and Bethune et al [2]. There is some controversy concerning the discovery of multiwall carbon nanotubes [3]. Some evidence suggests the discovery of multi walled carbon nanotubes in Russia [4] in 1952. Carbon nanotubes (CNT) are hollow nanotubules of carbon atoms with both partial alkene and aromatic characters. The nanotubule walls are honeycombs of fused hexagonal rings with nanoscale curvature and various symmetries: armchair, zigzag and chiral. The caps of the nanotubes contain hexagons and six pentagons with various arrangements depending on the size and symmetry of the tube. Carbon nanotubes have exceeded fullerenes as the most researched material of the $20^{th}$ century [3]. The CNT may be considered a close relative of fullerenes, even as elongated fullerenes, which were discovered in 1985 by Kroto, Smalley and Curl [5] and honored by the Nobel Prize in 1996. The CNT and fullerenes were discovered to form by the arc vaporization processes. Subsequently, CNT and fullerenes were also determined to be form by laser vaporization processes under appropriate conditions. Later, Richard E. Smalley also discovered that carbon nanotubes can be synthesized using a high temperature catalytic process (known as catalytic chemical vapor deposition (CCVD)) on certain transition metal nanoparticles, like Fe, Co and Ni and alloys thereof [6]. Micron particles of such transition metals (Fe, Co and Ni) were discovered many years ago to catalyzed the growth of micron-sized solid carbon filaments by Baker [7], as opposed to the hollow cylindrical tubes catalyzed by the nano-ferrocatalysts of Iijima [1] and Smalley [6]. In this work, the nanosize catalyst and resulting CNT is contrasted with the micron size catalyst and its catalysis for micron sized filamentous carbon for stressing the significance of important quantum and relativistic effects for this mesoscopic CNT new mechanism.

Baker [8] and Tibbett [9] have determined models for the formation mechanism of solid carbonaceous filaments on micron nanoparticles on the basis of classical transport effects of heat and mass, respectively. Prior explorations of CNT syntheses and mechanisms have attempted similar classical transport mechanisms with less success. Here it is introduced and determined that nonclassical dynamics of the mesoscopic catalyst lead to important nonclassical phenomena for CNT formation rather than filamentous carbon formation. The current experimental inability of controlling the diameter, helicity and number of sidewalls of CNT during synthesis is indisputable evidence of the mechanistic mystery and insufficiencies of the classical consideration prior to this work. In June 2000, Smalley stressed the incompatibility and the mystery of CNT formation mechanics [11]. In this work, experimental confirmation of a more successful model based upon a nonclassical dynamical ferromagnetic mechanism (proposed in July 2000) is presented [10]. This dynamical ferromagnetic mechanism presents new mesoscopic and electrodynamic phenomena not capture by older CNT mechanisms and filamentous carbon mechanisms. This search for a mechanism of CNT nucleation and growth was first inspired by the late Nobel Laureate Richard E. Smalley in June 2000 when he gave a keynote address to the Department of Defense Conference, "Out of the Box and Into the Future", at the Potomac Institute in Washington DC [11]. In this address, Smalley expressed the future technological importance and applications of CNT based upon their extraordinary properties. He also emphasized the unknown mechanism of CNT formation at that time and the need for greater mechanistic understanding for future technological applications.

After hearing Smalley's address, RB Little became immediately obsessed with explaining the CNT growth mechanism on the basis of the ferromagnetic character of the transition metals (Fe, Co and Ni) that best catalyze the formation with the simultaneous realization and explanation of transient ferromagnetism in both the catalyst and the growing CNT. Already studying the properties of ferromagnetic nanoparticles at the Naval Research Laboratory, RB Little formulated in July 2000 a ferromagnetic mechanism of CNT nucleation and growth and simultaneously discovered transient ferromagnetism in the growing carbon nanotubule. By developing a model based on ferromagnetism of the catalyst, ferromagnetic properties in the carbon were realized and given a physical basis. On the basis of the attributing transient ferromagnetism to carbon nanostructures, the effective catalytic ability of Fe, Co and Ni catalysts for CNT formation was explained. This dynamical ferromagnetic mechanism and the ferromagnetic basis of carbon were subsequently patented [10]. In the dynamical ferromagnetic model, a chemical and physical basis for ferromagnetism of carbon was determined and explained on the basis of strained, distorted, defective, pentagonal and heptagonal ring structures, and disordered covalent carbon bonds on the nano-length scale as carbon atoms agglomerate, nucleate and grow into defective, curvaceous carbonaceous nets and eventually into nanotubule structures in the ferromagnetic exchange field and under the organization of this magnetic field caused by the ferromagnetic catalytic nanoparticles. Such magnetism and polarization of the defective carbon bonds via strong exchange between carbon atoms, also causes strong exchange between the carbon and the metal for the organized, synchronized, stimulated, and coherent catalytic activity of the metal nanoparticle on over 100,000 carbon atoms per second. In arriving at this dynamical ferromagnetic mechanism, it was realized that such strained, distorted, defective and disordered covalent carbon structures are apart of the reactions coordinates during the CNT formation mechanism by which many carbon and metal atoms' intrinsic magnetism due to the altered bonds have enhanced exchange between these carbon atoms and the metal atoms (many bodies) of the magnetic catalyst nanoparticles for the organized, coherent, and synchronized catalytic activity on over 100,000 carbon atoms per second. Such spin orbital interactions between the carbon and metal result from Dirac nature of the spins, spin and orbital exchange between the carbon and metal atoms and quantum Hall type effects between the catalyst and surface defective graphene over the whole nanocatalyst that couples the whole growth processes: adsorption, decomposition, absorption, diffusion, graphene formation and CNT nucleation from the defective deforming graphene.

The Dirac nature of the spin currents and waves result from the nanosize of the system with many dangling metal and carbon bonds and the many body (spin) interactions, motion, correlations, and exchange. The resulting Dirac spins interact very intensely with the Landau levels in the catalysts for forming CNTs for the enhanced dense spin induced rehybridization (the Little Effect). The strong intrinsic dynamic magnetic field of the catalyst confines the relativistic Dirac spins. Such strong intrinsic dynamic magnetic field from the Dirac spins in the catalyst and relativistic orbital motion, dynamically organizes the pi bonds of the accumulating carbon atoms for the development of the ring currents and aromatization. The Dirac spin currents and spin waves are quantized for specific quanta of the dynamic magnetic field. The electrodynamic exchange between the spin currents and spin waves in the catalyst provides the exact quanta for accelerating the ring currents of the surface carbon into graphene rings. This quanta electrodynamics of the dynamic magnetic field of the catalyst and its acceleration of the ring currents in the forming graphene reflect a quantum Hall type interaction between the surface graphene and the ferrocatalyst. The changing magnetic field of the catalyst organizes the nonplanar ring structure in the forming graphene for distorted graphene and the nucleation of the CNT.

In this dynamical ferromagnetic model, many aspects of ferromagnetic carbon were first determined and revealed for the basis of this dynamical ferromagnetic catalytic activation of the CNT nucleation and growth. In addition to the novel aspects of the transient magnetic order in the carbon due to multiple bond rearrangement and defects on the nanoscale other phenomena such as surface tension, surface energy, shape effects and quantization of band structure at Fermi level were also reasoned for the realization of quite a distinct quantum dynamics and mechanics model of CNT formation relative to the older classical heat and mass transport models developed by Baker [8] and Tibbett [9] for filamentous carbon formation. The lack of accounting for novel mesoscopic phenomena was also identified for contributing to inadequacies of older models of CNT nucleation and growth.

In this mesoscopic, dynamical ferromagnetic model [10,12], many steps associated with CNT nucleation and growth that distinguished the CNT formation from the carbon filament growth, were introduced. The ferrometal catalyst through its interaction with the hydrocarbon effects a 12 step mechanism through the energetic, dense relativistic spin induced rehybridization, bond rearrangement, spin exchange and the catalytic magnetic quantum Hall type orbital acceleration for organizing, orchestrating and synchronizing many (over 100,000) carbon atoms per second into CNT. In particular, the comprehensive mechanism of CNT nucleation and growth involved at least 12 sequential, simultaneous and cooperative steps: 1. atomization of hydrocarbon by the catalytic decomposition by the ferro-nanocatalyst; 2. hydrogen and carbon adsorption, absorption, desorption and transport on, about and through the nanoferro-catalyst; 3. multi-spin induced orbital rehybridization of carbon and metal subshells; 4. carbon and hydrogen confinement within and on the nanocatalyst, 5. heat transport phenomena; 6. distortion and shape changes of the ferronanocatalyst-CNT composite; 7. carbon precipitation and clustering at a cooler side of the nanoferrocatalyst of higher spin density and ferromagnetism; 8. magnetic graphene growth from the carbon precipitate; 9. nucleation of ferromagnetic nanotubular structures as the graphene grows; 10. dynamic ferromagnetic driven growth of the CNT; 11. growth and reconstruction of the ferromagnetic catalytic nanoparticle; 12. poisoning of the ferromagnetic nanocatalyst particle. The essence of this model is that nonclassical dynamical ferromagnetic phenomena, of spin waves and spin currents on the nanoscale activate, organize, stimulate and synchronize the mechanics of carbon, hydrogen and metal atoms involved in these 12 steps. On this basis, the catalyst can organize over 100,000 carbon atoms per second on the basis of this dynamical ferromagnetic mechanism for the high efficiency required for the observed rapid growth rates of over 1 micron per second. There must be some organizing phenomenon in order for the catalyst to process so many delicate unstable carbon atoms and bonds over an extended time. The dynamic ferromagnetism is the organizing factor. Pure coulomb interactions cannot account for the observed rapid simultaneous organization of so many carbon atoms and their bonds. Without such dynamic ferromagnetic organization highly metastable intermediates of metal carbides, hydrocarbons, and amorphous carbon would form rather than the crystalline CNT side wall.

Unlike the older CNT and filamentous carbon growth models, this model includes the quantum effects of the physicochemistry of the electrodynamics of carbon and metal atoms rehybridization and the chemistry of carbon, hydrogen and metal bond rearrangement on the mesoscopic scale with consequent non-classical and novel effects. Such a novel environment leads to novel chemical dynamics associated with dense Dirac spins (their greater effects due to the nanosize, surface dangling bonds and surface energy) and the effects of these Dirac spins on multi-atomic orbital rearrangements (Little Effect). Novel chemistry results from high energetic bond rearrangement and extensive organizing magnetic exchange by these Dirac spin and Landau currents for quantum Hall type acceleration of distorted ring currents to form graphene. Indeed, recent research has determined Dirac spins in magnetic transition metal nanoparticles [13] and also Dirac spins in nanographene in strong magnetic field [14]. It is important to note the forming surface graphene compresses the ferro-nanocatalyst by the distortion step of the mechanism for greater Dirac nature of spins even in the metal ferrocatalyst.

The dynamical ferromagnetic CNT mechanism had previously put the two (ferromagnetism of catalyst and forming CNT) together in July 2000 [10]. These novel spin and organized energetics result in novel chemical bond rearrangement on the basis of the Little Effect. Such rehybridization and bond rearrangement chemistry are essential for many of the 12 steps of CNT formation: hydrocarbon decomposition, carbon uptake, hydrogen uptake, carbon and hydrogen transport around and through the catalyst, catalytic distortion and the graphene and CNT nucleation and growth. This mechanism outlines the concerted, synchronized, cooperative and stimulated nature of these various processes. Moreover, the mechanism introduced dense relativistic spins, currents and magnon phenomena for organizing such multi-atomic chemical dynamics for the quantum mechanism. This newer mechanism accounts for the quantization of these spins, currents and magnons in their catalytic roles for forming CNT. Such quantization of frontier states of the catalyst explains the resulting possible symmetries of the CNT, and the difficulty of controlling the CNT growth without controlling the dynamical spin currents and spin waves in the catalyst. On this basis, this dynamical ferromagnetic mechanism accounts for various processes of CNT nucleation and growth in terms of a dynamical ferromagnetic environment causing multi-spin orchestrated orbital dynamics for concerted, synchronized, cooperative and coherent carbon decomposition in a hot zone, carbon transport to a cooler zone, carbon rehybridization and carbon bonding to graphene structures in a cooler zone and subsequent graphene growth and distortion. On the basis of this dynamical ferromagnetic model, dynamical fields should be able to drive, control and analyze the CNT mechanism of formation.  Therefore, this dynamical magnetic mechanism provides a better control for better synthesis.

In this paper, we present experimental evidence and support for this nonclasical dynamical ferromagnetic CNT formation mechanism by imposing dynamical electromagnetic fields to observe and thereby prove effects on the process and the appropriateness of this dynamical ferromagnetic field model.  First, prior experiments associated with catalytic chemical vapor deposition (CCVD) with external furnace (EF) heating in strong static external magnetic fields up to 20 teslas are considered.  Moreover, a recent RF - CCVD process by Biris et al., [15] for dynamically modulating, electromagnetical activation and organization of magnons and spin currents in the catalytic ferronanoparticle via RF radiation are presented and compared with the CCVD in static and zero external magnetic fields for direct evidence of the nonclassical dynamical ferromagnetic CNT growth mechanism.

**Experimental Procedure**
The experimental procedure involved doing catalytic chemical vapor deposition of carbon nanotubes within different magnetic environments and using two types of catalysts.  The carbon nanotubes formed by these various methods were analyzed by various techniques.  First, the catalyst preparations are presented.  The two catalytic CVD heating systems are then presented and finally the analytic methods used on the CNT are discussed.

Several catalytic systems have been used to grow both single and multi wall carbon nanotubes. For this experimental study, a Fe-Co/$CaCO_3$ catalyst was used to grow the MWCNT, which was prepared as previously explained by Couteau et al [16]. The stoichiometric composition of the catalyst was $(Fe_{2.5}Co_{2.5}):(CaCO_3)_{95}$.  Multiwall carbon nanotubes were synthesized using acetylene as carbon source by a CCVD approach [17]. Once the temperature reached 720 $^o$C, the acetylene at 4.3 ml/min was introduced inside the reactor for about 30 minutes.  Neither growth of nanotubes nor any other types of carbon byproducts was found in the experiments performed only with a graphite susceptor without catalyst.  Single wall carbon nanotubes were grown at 850 $^o$C on a Fe-Mo/MgO (1:01:12) with methane as a carbon source and Ar as the carrier gas.

For the synthesis reactions, around 200 mg of catalyst was uniformly spread into a thin layer on a graphite susceptor and placed in the center of a quartz tube positioned horizontally inside an external furnace (EF) and inside an inductive furnace under nitrogen flow at 300 ml/min. Radio frequency (RF) excitation (350 kHz and 1.2MHz) was used to energize the graphite susceptor that contained the catalyst [15].  The as-produced nanotubes were purified in one easy step using diluted hydrochloric acid solution (1:1) with sonication for 30 minutes, followed by distilled water washing and dried at 120 $^o$C overnight.

To understand the relationship between the catalyst activation via external conductance heating (EF) versus external RF activation and carbon nanotube growth, the catalyst and CNTs were characterized by using scanning electron microscopy (SEM) transmission electron microscopy (TEM), thermogravimetric analysis (TGA), and Raman scattering spectroscopy. The morphology of catalysts was monitored with a JEOL 6400F high-resolution scanning electron microscope. TEM pictures of the carbon nanotubes were obtained from FEI Tecnai F30 transmission electron microscope. Thermogravimetric analysis was used to study the thermal behavior of the catalyst system and also to determine the overall purity of CNTs. Thermogravimetric analysis was performed under airflow at 150ml/min using Mettler Toledo TGA/SDTA 851e and was used to monitor the weight losses and the thermal stabilities of the samples. Raman scattering studies of the catalyst and CNTs were performed under room temperature using Horiba Jobin Yvon LabRam HR800 equipped with a charge-coupled detector, a spectrometer with 1800 and 600 lines/mm gratings and He-Ne (633 nm) and $Ar^+$ (514 nm) lasers as excitation sources. The microscope, a confocal Olympus microscope (high stability BX41) equipped with Olympus objectives (100x, 50x, 10x), focused the incident beam to a spot size of <1 $\mu m^2$ and the backscattered light was collected 180° from the direction of incidence. Raman shifts were calibrated with a silicon wafer at the peak of 521 $cm^{-1}$. The spectral resolution was 1 $cm^{-1}$ and the collected signal was averaged over 10 cycles.

**Experimental Results and Discussions**

The experimental procedure was kept uniform during the CNT syntheses (catalyst, reaction temperature, carbon source and carrier gas, flow rates, reaction time), and the only parameter that was changed was the energizing mechanism: RF or EF energizing. The RF oven is shown in Figure 1. Compared with the regular CCVD, involving external furnace (EF) heating (where the heat comes from the outside of the reactor with consequent gradient in temperature over the catalyst surface), the RF heating process, involves induction heating of the graphite susceptor that supports the catalyst as well as energizing the metal nanoclusters on the graphite susceptor for activating the catalytic system. Whereas the EF process more stochastically introduces energy into the ferronano-catalyst, the RF process more orderly deposits energy into the ferronanocatalyst with fewer proclivities to disorder and discord. Therefore, it is expected that the RF excitation of the metal clusters would affect the reaction kinetics and morphology of the nanotubes in a different and better way than the conventional EF heating technique due to the nonstoichastic organization of spins and spin currents by the electromagnetic radiation upon the magnetic nano-metal clusters by the RF radiation. The nanosize of the metal clusters is demonstrated to enhance the spin current and spin wave effects beyond the heat effects due to the larger level spacing and the slowing of the relaxation of the magnons to heat.

As a result of using these different energetic inputs, quite different CNT properties were observed. Figure 2 shows the lower magnification TEM pictures along with the diameter variation plots for the multiwall carbon nanotubes grown in EF and RF (at a 350 kHz frequency) systems. It can be observed that a shift occurs towards smaller values of the outside diameters for the multi wall carbon nanotubes grown under RF energizing as compared to those grown in EF heating. The higher magnification TEM pictures shown in Figure 3, indicate that the nanotubes grown under RF radiation have a smaller number of graphitic walls and larger lumen and that as the RF radiation frequency increases the CNT diameter decreases (Figure 4). The same size trends were observed for both the multi and the single wall carbon nanotubes. Another interesting experimental observation was the higher degree of crystallinity for the nanotubes grown under RF. See Figures 5, 6 and 7. Raman Spectroscopy is a widely used method to evaluate the crystalline properties of both the single wall and multi wall carbon nanotubes. The Raman spectra for such materials have several characteristic bands. Radial Breathing Mode (RBM) between 100 and 400 $cm^{-1}$ is directly related to the diameter of the single wall nanotubes and is absent for multiwalls. The D band (around 1300- 1350 $cm^{-1}$) corresponds to the defects in the nanotube walls, while the G band (between 1500 to 1580 $cm^{-1}$) in given by the in-plane stretching of the carbon-carbon bonds. An overtone of the D band (2D), found between 2600 and 2650 $cm^{-1}$ is also an expression of the crystallinity of the nanotubes. The ratio between the intensity of the G and D bands (I(G)/I(D)) is a widely used measure for the crystallinity for the nanotubes. The higher this ratio, the more crystalline the graphitic walls are [18]. Figure 5, shows the variation of the I(G)/I(D) for both single wall and multiwall nanotubes grown in regular CCVD and under RF excitation with a frequency of 350 kHz, indicating an increase in the crystallinity of the nanotubes grown under RF activation. The same conclusion can be drawn also from Figure 6, which shows the DTA curves for the multiwall carbon nanotubes grown by EF and RF processes. Figure 7 gives a lot of data concerning mass percentages of amorphous carbon in the CNT products of different growth periods by the EF and the RF methods. The RF method generates much less amorphous carbon over longer growth times relative to the high and increasing amorphous carbon amounts observed in the EF process. Generally, the lack of defects in the graphitic walls indicates a higher temperature of thermal disintegration during DTA.

On the basis of these differences in CNT size, number of sidewalls, size distribution, crystallinity and purity observed from RF and EF methods evidence of the validity of the nonclassical dynamical ferromagnetic mechanism of CNT formation is provided and demonstrated. Here, the range of possibilities (static, dynamic, and zero applied field) of magnetization of the catalyst (externally influencing its spin currents and spin waves) for testing this dynamical magnetic CNT mechanism is considered. The resulting data reflect the extent of experimental possibilities of magnetic environments of the catalyst: 1. zero applied, terrestrial magnetic field, 2. strong external static applied magnetic fields, 3. intensely applied, dynamical, magnetic fields of radio frequency electromagnetic radiation. The results from this range of experimental external magnetic probes of the mechanism coincide with the proposed dynamical ferromagnetic mechanism in terms of the transient magnetic order associated with spin waves and spin currents causing the proposed 12 steps of the mechanism of CNT formation. On the basis of the dynamical ferromagnetic mechanism of CNT nucleation and growth [10,12], under conventional EF synthesis conditions the ferro-catalytic nanoparticles use the input heat from conduction to stochastically develop their own intrinsic spin currents, magnons and spin waves. Also in the conventional stochastic EF method, there are many random energetic disturbances. It is important to note that the EF heat does assist the breakage of chemical bonds and provides the energy to develop the spin currents and waves although rather inefficiently. But the bonds are not broken entirely by the thermal effects of the heat, the electron currents and polarized spins of catalysts are actively breaking the bonds and organizing new bond formation. The complete thermal decomposition would require much more heat over 4000 $^{o}$C. The electrons of the catalyst break the bonds of carbon with less needed heat. Electrons of the best catalysts Fe, Co, Ni are transiently spin polarized during the process. Other transition metals are less catalytic for CNT because they have less polarized spins. It is the polarized electrons of the catalyst (by Little Effect) that catalyze bond breakage and reformation during the 12 steps. An external magnetic field can drive these polarized electrons, magnons thereby driving the carbon via the spin polarized electrons of these catalytic metals (Fe, Co and Ni). It is an experimental fact of many other researchers that the external magnetic field can orient the polarized electrons of these metals. Electrons of these metals may be exposed 45 Tesla external magnetic fields via electromagnets to control the chemistry. Moreover, here it is determined that the intrinsic even much higher magnetic fields of several hundred tesla via coupling to the ferrometals (Fe, Co and Ni) even under synthesis conditions cause the formation of the CNT.

Such fruitful self-organizing spin current and spin wave dynamics activate and organize the dynamical electromagnetic processes of carbon, hydrogen and metals atoms during the 12 steps of the mechanism. As explored here, external magnetic fields will either perturb these intrinsic spin and current dynamics to retard the CNT formation or enhance and drive these fruitful intrinsic spin dynamics in a constructive way to enhance the growth of the magnetic CNT. Other investigators [19] have determined that the RF radiation excites magnons and can change the anisotropy of the pure ferromagnetic nanoparticles. Therefore, RF radiation does excite spin waves and spin currents in the ferro-nanocatalyst. Here, the effects of such RF driven magnons and spin currents within the nanocatalyst are presented for organizing CNT formation, thereby substantiating the proposed dynamical ferromagnetic mechanism. It is important to note that the magnons driven by the RF are more important than the heat from the RF due to nanosize and larger level spacing due to the nanosize. This quantum size effect (QSE) differentiates the CNT growth mechanism from the filamentous carbon growth mechanism. The RF radiation does produce heat in the nanoparticle but the degeneration of the magnon to heat in the nanoparticle is slower than the degeneration of magnons to heat in the larger particles of micron sizes due to lower level spacing in the larger micron size particles. This causes less heat effects in the nanocatalyst and less heat effects during the CNT nucleation and growth. This is another reason for this new mesoscale dynamical magnetic model, which accounts for these effects not considered in the Baker and Tibbett models of filamentous carbon formation. It is important to note that such intrinsic or externally driven magnons and spin currents during the CCVD processes on the basis of this model are in analog to electrons and their vortices in the plasma in the electric arc or laser vaporization processes [12]. These magnons and spin currents act as Dirac spins and Landau states for the synchronous, orchestrated and stimulated transformations of the bonding and electromagnetic states of carbon, hydrogen and metals atoms within these systems to cause the 12 step dynamical ferromagnetic mechanism of decomposition, absorption, rehybridization, transport, structural dynamics, carbon clustering and CNT nucleation and growth. The magnons and spin currents thereby drive and organize the formation of the magnetic carbon nanotube intermediates and with the eventual annealing into the CNT final Luttinger state by organizing the carbon atoms and their spins and orbital motions for bonding and forming magnetic intermediates that relax into the less magnetic CNT product. The forming and growing magnetic CNT (defective, distorted nanographene) exhibits similar magnon and spin current excitations as in the catalytic ferromagnetic nanoparticles during the catalysis. The magnons in the catalysts template and map the magnons in the forming CNT by organizing the CNT's size and helicity. External magnetic fields (dynamics or static) can confine these Dirac spins and orient the forming Landau levels. Both recent theoretical evidence [20] and experimental evidence [21] of such ferromagnetism in nanographene as predicted in [10] have been reported in the literature in 2005 and 2006. Thereby the electromagnetic impulses that excite the ferro-nanocatalyst also organize the formation, couples to and excite the CNT and its synthetic intermediates for the naturally stimulated and organized the formation of these important intermediary states along reaction coordinates to catalyze the CNT formation. The observed effect of external RF radiation presented here on size control, number of sidewalls, growth time, CNT length crystallinity and purity are direct proof of this. Many of these intermediary states have very short lifetimes, but other researchers have experimentally demonstrated ferromagnetic ordering in the parts (the catalytic ferro-nanoparticle and the nano-carbon) on femtosecond time scales [22]. These aspects of the intrinsic magnons and spin currents of the ferronanocatalysts and the external electromagnetically driven modes of these magnons and spin currents are here presented and experimentally demonstrated to cause the formation of the CNT.

The observed differences in formations of CNT in strong external magnetic fields of various spatial and temporal patterns are evidence of this significant intrinsic dynamic magnetic factor during the carbon nanotube formation. This reflects how the different magnetic environments influence the spin currents and spin waves that organize the CNT nucleation and growth. Whereas a strong static magnetic field disrupts the mechanism of CNT formation [23,24], a strong dynamic magnetically driving environment via the RF excitation enhances the CNT formation even such that the resulting CNTs form more uniform distributions (See Figures 2 and 3) relative to CNTs formed in only the (zero applied) terrestrial magnetic field during the conventional EF method. See Figure 2a. The conventional EF process forms broader CNT size distributions (See Figure 2a) because during this process each nano-catalyst stochastically develops its own internal spin currents and magnons that form a variety of CNT sizes on the basis of the original catalyst size distribution. The stochastic heating by the EF process also subjects the self organized magnons and spin currents in the catalyst to more non-resonant disturbances that can poison the process, shorten the growth time and contributes to amorphous carbon formation. See Figure 7. But the RF excitation provides an ordered external electromagnetic organizing force for directly generating specific spin current and magnon modes in the nanocatalyst (nonstochastically) for forming CNT of more specific sizes for more narrow size distributions. See Figure 2b. Due to nanosize the RF excited magnons degenerate more slowly to heat for better use of the magnons to form CNT. Furthermore, the greater organization by the RF process presents fewer environmental disturbances for more prolong growth relative to the CNTs formed by the perturbable EF process and with greater crystallinity by the RF process. The specific frequency of the RF radiation organizes magnons and spin currents in the catalysts for organized templation and catalysis of CNT of specific sizes on the basis of the dynamical ferromagnetic mechanism. The importance of such size effects by the RF method involving a single wavelength to template and catalyze the CNT formation is revealed in Figure 2b, wherein the size distribution shows a selection of tubes of sizes: 17.5 nm, 35 nm and 52 nm. This corresponds to a period in CNT size of about 17.5 nm. This is consistent with the dynamical ferromagnetic mechanism on the basis that a specific frequency of the RF radiation corresponds to specific natural magnon and spin current frequencies in the nanocatalyst of specific sizes for catalytic activity of specific spin waves and currents for inducing CNTs of specific sizes and orbitals and spin wave quanta. The nanocatalysts may even under the specific wavelength of the RF radiation and synthesis conditions alter their aspect ratios for more efficiently nucleating and growing CNTs of matching, favored sizes and with the magic number radii. Such size selection and magic number radii are proof of the proposed mesoscopic quantum phenomena proposed in the dynamical ferromagnetic CNT mechanism. Structural and shape alterations of the catalyst during the CNT growth are normally observed in accord with step 6 of the mechanism [12]. The RF radiation may select specific size (size effect, quantum effect) nanoparticles of compatibility with its wavelength for shorter induction time and faster CNT growth.

These results and there explanations by the dynamical ferromagnetic mechanism are consistent with Iijima's [25] recent discovery of RF plasma synthesis of MWCNT of specific size without transition metal catalysts (Fe, Co and Ni). Iijima 's [25] process is autocatalytic and as predicted by this ferromagnetic dynamical model the carbon forms clusters that are themselves ferromagnetic and act magnetically in analog to the Fe, Co and Ni nanocatalyst (considered here) to autocatalyze the CNT nucleation and growth on the basis of natural characteristic spin waves and currents within the carbon nanocatalyst itself driven by the external RF radiation to autocatalyze its own graphitic nanotube formation with size selection by the RF radiation. These results of Iijima together with the results presented here give evidence and beauty to the dynamical ferromagnetic mechanism and the adjoining ferromagnetic carbon disclosed in the mechanism [10].

It is very interesting to provide continuity between this CNT mechanism and the mechanism of micro-filamentous carbon formation. Unlike the nanocatalyst discussed here with its mesoscopic size and discontinuum of electronic states for quantum mechanical description and the dynamical ferromagnetic mechanism, the micron catalyst used in forming carbon filaments have more continuum of states. The continuum of electronic states contribute to more continuum of spin waves and spin currents within these bigger micron catalyst for less electronic restrictions and surface effects in comparison to the discontinuum of levels in nanocatalyst and CNT process. The greater density of magnon states allows greater role of heat in the process of forming filamentous carbon and greater classical description of the filamentous process. The greater density of states in the larger micron catalyst for carbon filaments cause lower energy Dirac spins, less quantization, greater degeneracy rates of magnons to heat for more planar graphene formation. This leads to the more efficient and even reversible catalyze micron carbon filament nucleation kinetics as demonstrated by Baker [26]. The CNT growth is less reversible by EF processes. The graphitic planar micron sheets nucleate faster for the micron catalyst but such planar sheets are kinetically restricted on the nanoparticle catalysts with thermodynamic instability of the planar graphene, so the planar structures distort into hemispheres and elongate into nanotubules with faster kinetics relative to planar stacking of nano-graphenes. The planar nucleation requires more magnetic anisotropy in the catalyst, the micron catalyst exhibits greater magnetic anisotropy for causing faster planar graphitic nucleation and growth.

On the other hand, smaller catalytic ferroparticles on the nanoscale have greater surface to volume with greater surface induced spin waves, smaller coercivity, magnetization reversal modes, and smaller dynamic magnetic susceptibility [27]. Here it is suggested that the curvature and surface effects in the ferro-nano transition metal catalyst can contribute to complex surface saturation, unsaturation and bonding dynamics by delta bonding surface d orbitals for novel ring currents, spin currents for novel surface saturation and dangling bond capacities for explaining the compatibility of 3d nanoparticle surfaces to pi bonding and aromatic dynamics of carbon p orbitals to organize tubular nanographene. On the basis of the smaller nanosize, the dangling bonds and surface energy cause greater relativistic effects of the Dirac spins, greater quantum effects, quantum Hall type acceleration of ring currents for nonplanar graphene under the extremely dynamic magnetic field of the catalyst, and more difficult dissipation of magnons to heat, these differences cause CNT catalysis for nanosize catalyst as opposed to the planar graphene for the larger micron catalysts. The different magnetic properties of micron catalyst vs. nanocatalyst cause slower kinetics of planar nanographite formation and faster kinetics of nonplanar nanotubular structural formation. The nanocatalysts tend to be single domain which couple well with recently confirmed ferromagnetism in nanographene [28], distorted graphene [29] and more over contact induced ferromagnetism in graphene and CNT [30]. The quantum description of the nanoparticle mechanism expressed here requires more specific energetic quanta, spin waves and spin currents within the mesoscopic nano-catalysts, which cause slower planar graphitization kinetics, and faster nonplanar and tubular graphene kinetics. The spacing in the nanosize results in the lesser degeneration of the spin currents and waves to heat in the nanocatalyst for better organized motion and spin currents and waves for organizing the distortion of the graphene into CNT and growth of the CNT. It is important to note that this RF method provides less stochastic excitation in the mesoscopic nanocatalysts for greater efficiency of these characteristic energetics, spin current and magnon modes for shorter induction time, faster CNT growth, longer CNT growth and as discussed below more efficient use of carbon during the CNT nucleation and growth for greater crystallinity and less amorphous carbon formation. The external RF field can be tuned to these spin currents and spin waves to indirectly organize and control the CNT nucleation and growth.

Even more evidence of this ability of external dynamics magnetic field to organize specific sizes of CNT is the observed reduction in the number of sidewalls during the RF excited CCVD of CNT relative to the EF- CCVD process. See Figures 3 and 4. The greater organization of the characteristic natural frequencies and the large separation of these natural frequencies of spin currents and magnons of the catalyst by the external RF radiation is shown to reduce the number of walls thereby further attesting to the electromagnetic ability to control nucleation and enhanced growth of CNT of specific diameters by matching the catalyst and CNT sizes with the frequency of the RF radiation. Just as the RF radiation organizes spin-induced currents for catalyzing CNT formation of specific radii, it hinders formation of concentric shells having unfavored radii. The magnons in the forming CNT must match the magnons in the catalyst, which is tuned by the RF radiation. These observations are evidence of the validity of the intrinsic dynamic ferromagnetic nature of the CNT formation mechanism and the ability of the external electromagnetic radiation to drive specific internal spin processes. Such size specificity of radio frequency radiation further limits multishell development due to the resonant electromagnetic RF driving natural modes of magnon, spins and ring current dynamics in the catalyst. These natural catalytic modes favor electromagnetically organized nucleation of CNT of specific size and hinders other sidewalls of different sizes and different orbital chirality, size and spin modes. The nucleation of new sidewalls requires new cluster formation, cluster growth, CNT nucleation of different sizes and CNT growth of different size of incompatibility to the driving RF frequency. But the growth of existing CNT of the favored radii occurs faster than the nucleation of new CNTs of unfavored radii. Therefore the RF radiation slows new nucleation for formation of CNT with fewer walls. The EF method does not so limit the intrinsic magnons and ring currents to specific sizes for nucleating specific size CNT. So the size distribution is greater for the EF method. Due to the nanosize effect there is a need for a new model that captures these quantized spin wave, spin current effects on the nanoscale for the unique dynamics of the CNT relative to the carbon filament formation process.

Such specificity of the size by the electromagnetic field of the RF radiation for the softening and distortion of the ferro-catalyst for specific diameters by altered aspect ratio of the ferrocatalyst is further demonstrated by the observed frequency dependence of the ferrocatalyst size and ferromagnetic CNT size. As the frequency of the electromagnetic radiation increases, the size of the magnetic CNT decreases. See Figure 2. This decrease in CNT size as the RF radiation frequency increases is a result of the higher frequency causing higher frequency magnons, spin currents and skin currents in the catalysts for their organization of smaller magnetic CNT diameters with more compatible magnons and spin and orbital currents by alteration of aspect ratio of the catalysts. The faster electromagnetic modes in the ferromagnetic catalyst (excited by high frequency radiation) organize nucleation, growth and couple to the smaller ferromagnetic intermediary CNT, having corresponding comparable energetic, spin waves and spin currents with the high frequencies in the catalyst and in the higher applied RF radiation. In addition to such size control, here, it is suggested that some combination of external dynamical magnetic and/or electric fields may select CNTs of specific helicity.

The carbon nanotube itself is ferromagnetic and the disordered, defective, distorted covalent bonded intermediary states that form along the reaction trajectories of CNT formation lead to the ferromagnetism of the CNT and its efficient coupling to the transient ferromagnetic states, spin currents and spin waves within the catalyst during the process and mechanism. On this basis, the electronic states of various CNT intermediates are subject in fruitful ways to external magnetic fields of the catalyst and even reinforcing external dynamic radiation fields as thereby the RF radiation. Such transient ferromagnetism of carbon readily allows it to couple electromagnetically such that intrinsic and externally, electromagnetically driven spin and charge dynamics within the catalyst couple with the carbon to organize its growth. The ferromagnetism of the CNT intermediary states is consistent with the observed ferromagnetism of the catalysts particles after CNT growth [31] and the high saturation magnetization of Fe, Co and Ni filled CNT [32] are evidence of this ferromagnetism of carbon during the growth. Furthermore other investigators have observed the transient ferromagnetism in the parts for this complex system. It was observed [22] the transient ferromagnetism in ferrocatalytic nanoparticles on femtosecond time scales and [33] observed ferromagnetism in disordered carbon.

This role of the intrinsic and reinforcing externally driving electromagnetic field for affecting the ferrocatalytic size by distortion and promoting ferromagnetic CNT construction of specific size and having specific magnon and ring current properties is further supported by the observed greater crystallinity of the ferromagnetic CNT formed by the RF process relative to the EF process. See Figures 5 and 6. Thus the dynamic ferromagnetic field by the external RF not only organizes the mesoscopic aspects of clustering and CNT nucleation, the external RF radiation also organizes the earlier steps in the process of atomization, atomic rehybridization, transport and C-C bonding. The RF method more directly organizes the needed spin and orbital currents for the carbon rehybridization and rebonding in comparison to the more random intrinsically, more easily perturbed spin and orbital dynamics developed within the catalysts in the EF process. The greater electromagnetic organization, synchronization and stimulation by the RF process leads to better organization and symmetry of the chemical bonds in the graphene products. It is demonstrated here that the dynamic ferromagnetism of the catalyst acts on the nucleated graphene such that it affects a dynamical quantum Hall effect for accelerating different directional ring currents in the graphene for causing the curvature of the graphene and its transformation into the CNT. Such quantum Hall effects by bigger micron catalysts accelerate ring currents in the same plane for filamentous carbon formation rather than CNT formation. Such quantum Hall effects for accelerating ring currents in the graphene cause greater organization of the bonding in the CNT and graphene products for their greater crystallinity. Therefore the CNT product from the RF method on the basis of the ferromagnetic mechanism should have greater crystallinity than the products from the EF process. Since the model determines that it is the intrinsic magnons and ring currents in the ferrocatalyst nanoparticles that organize and direct carbon transport, surface reconstruction and carbon chemistry for clustering, graphitization and growth, the quantum Hall like interactions between the catalyst and the forming CNT diminish amorphous carbon deposits. The spin currents and magnons cause the graphitization and minimize amorphous carbon formation. Since the external RF radiation drives such fruitful surface magnons and spin currents, the RF radiation should more efficiently organize carbon bonds for greater crystallinity.

Even more evidence for the greater electromagnetic organization of carbon bonds of the RF method is revealed by the DTA curves (See Figures 7 and 8) which demonstrate that the RF process generates less amorphous carbon and more crystalline carbon in CNT due to the RF radiation's greater organization of the carbon bond rearrangement to CNT. The observed greater buildup of amorphous carbon during the EF process as a function of time is more evidence that the intrinsic magnetism, magnons and spin currents play important roles during the growth mechanism. As carbon builds up in the catalyst of the EF process the ferromagnetism diminishes with the eventual poisoning of the catalysts as outlined in step 12 in the dynamical CNT ferromagnetic mechanism [10]. RF generates less amorphous carbon due the direct excitation of the magnons in the nanocatalysts and their slower degeneration to heat for more direct activation of the catalytic process and better use of carbon atoms to form the crystalline CNT structures. It is important to note the catalytic activity is not simply a charge transfer but a spin polarized charge transfer between the catalyst and the carbon for the catalyzed formation of the CNT. Many transition metals can affect a charge transfer but only Fe, Co and Ni under the observed conditions can affect spin polarized charge transfer. RF radiation directly couples to the spin polarized charge in these catalysts. The EF heating requires that the heat organize polarons and magnons, which is much less efficient than the direct RF excitation. On the basis of this new mechanism, the older classical mass transfer and heat transfer mechanisms used for the micron or bigger catalysts are not sufficient. Quantized spin current and spin wave transfer through the catalyst are necessary as put forth in this new dynamical magnetic mechanism. Such magnetic transfer is the basis for the efficient organized transport and gathering of over 100,000 carbon atoms per second into CNT.

**Conclusion**

Catalytic chemical vapor deposition (CCVD) process is the most promising method for producing larger quantities of carbon nanotubes with controlled morphologies, since it allows a more direct control of the physico-chemical parameters that affect the synthesis process, such as: temperature, carbon sources, gases flow rates, catalytic systems. Although significant advancements were made during the last several years, still the growth process of carbon nanotubes is under significant investigation given its complexity and the large number of parameters that influence it. One of the most advanced models that explain the formation of carbon nanotubes was developed by R. B. Little and consists of twelve stages, as previously shown. This model shows the strong correlation between the electron spins of the carbon and metal atoms in the catalyst and the bond rearrangement dynamics for CNT formation about the catalyst. To further prove this model, various types of carbon nanotubes (both single and multi walled) were synthesized by CCVD in an external furnace (EF) and by RF excitation at two frequencies of 350 kHz and 1.2 MHz. It was demonstrated that the nanotubes grown under RF excitation show smaller diameter and fewer numbers of walls and show a higher degree of crystallinity with lesser amounts of non-crystalline (amorphous) carbon as compared to those synthesized in a regular furnace by EF heating. These findings can be explained by the interactions between the external electromagnetic fields and the magnons and spin currents in the ferrocatalyst nanoparticles, which organize and direct carbon transport, surface reconstruction and carbon chemistry for clustering, graphitization and growth of carbon nanotubes. The RF activated CNT formation even demonstrated the selective formation of CNTs of specific sizes on the basis of the RF frequency and resonant excitation of the catalysts. But unlike the dynamic electromagnetic field for resonating fruitful magnons and spin currents by the RF irradiation of the catalyst, the CCVD within the bore of a strong DC magnet for strong static magnetization of the catalyst retarded the growth of the CNT. These results provide explicit proof of the mesoscopic, dynamical nonclassical ferromagnetic mechanism of CNT nucleation and growth. Recent observations of ferromagnetism of carbon, high temperature quantum Hall effects in graphene and relativistic Dirac spins in graphene and ferrometals support the prior dense spin orbital dynamical model of RB Little.

**Acknowledgement**
With external gratitude to GOD - RBL.

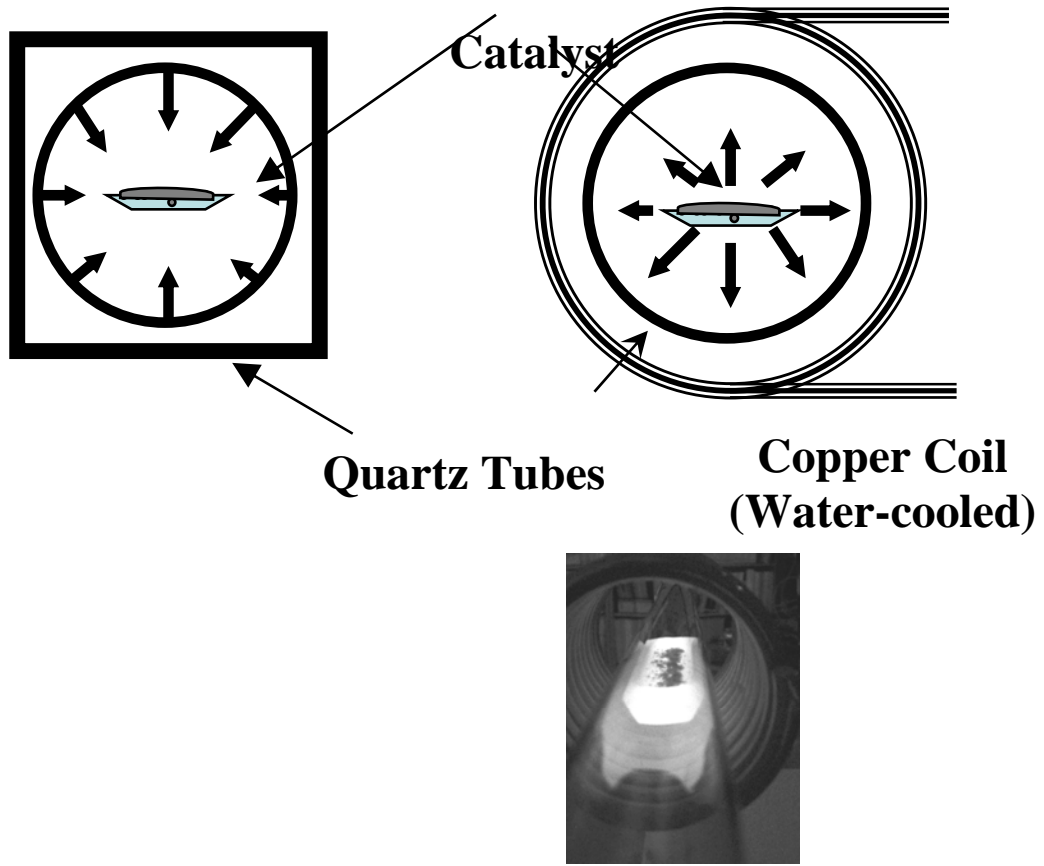

Figure 1. The schematics for the resistive heating with external furnace (EF) and the inductive heating (RF) processes and the RF Reactor during the carbon nanotubes growth reaction. The catalyst system is placed in the quartz reactor and heated by inductive currents generated by the induction coil. The electromagnetic field induced by the RF coil induces significant modifications in the metallic catalyst systems, which can be reflected in nanotubes with altered morphologies and properties.

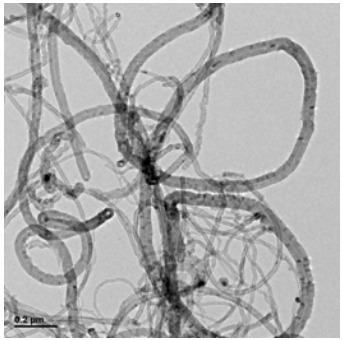
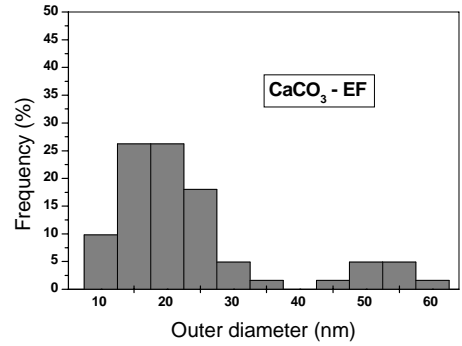
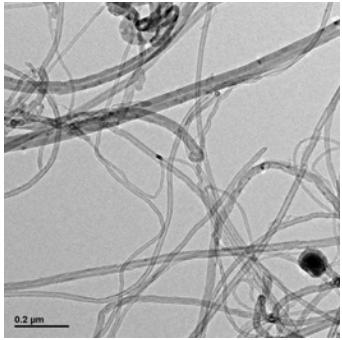
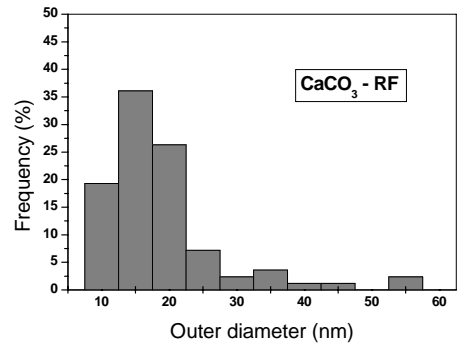

Figure 2. The low-magnification TEM images and the statistic analysis of diameter distribution of the nanotubes synthesized by the two different methods a) outside furnace, and b) RF oven. [Biris, A. R.; Biris, A. S.; Lupu, D.; Trigwell, S.; Dervishi, E.; Rahman, Z.; Marginean, P., *Chem. Phys. Lett.* **2006**, 429(1-3), 204]

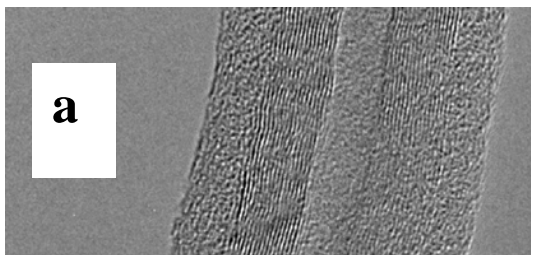

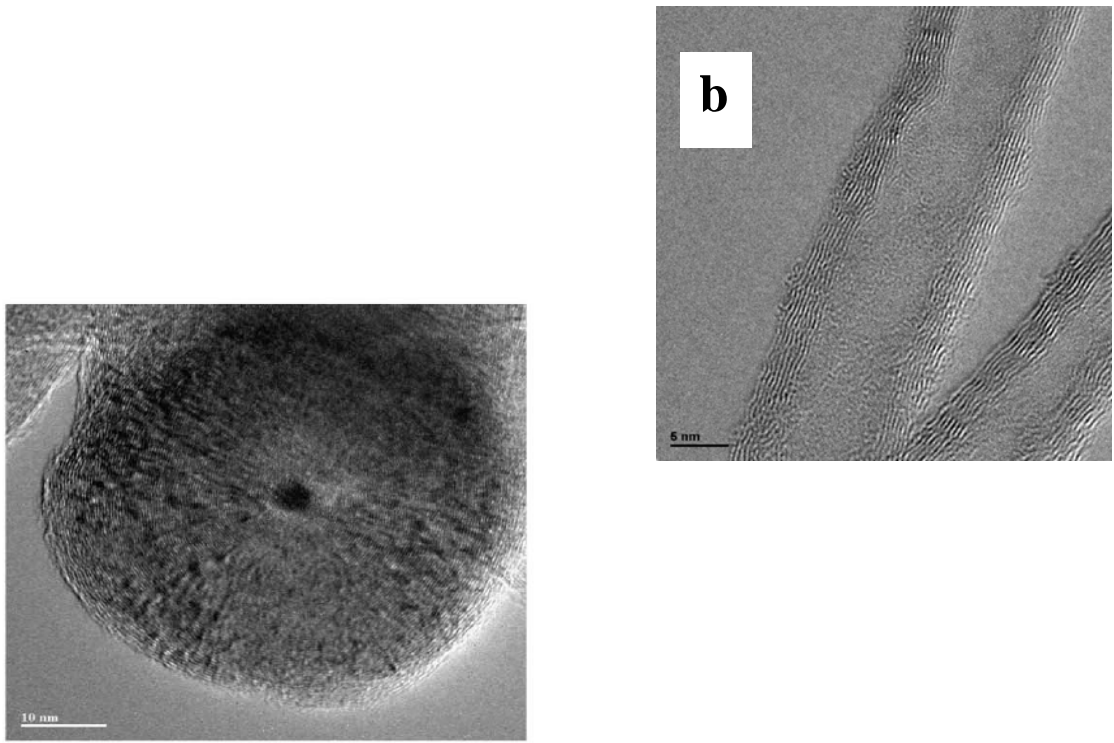

Figure 3. The high magnification TEM images of the nanotubes produced by two different methods (a) outside furnace and (b) RF after 50 minutes of growth reaction. It can be observed that the nanotubes produced by regular thermal furnace heating have a larger number of carbon layers and are covered by a layer of amorphous carbon, while those grown in RF do not visually show any amorphous carbon. The TEM picture taken at the end of the nanotube grown by regular thermal heating show a small lumen with a large number of carbon layers. The lumen is much larger for the nanotubes grown in the RF oven.

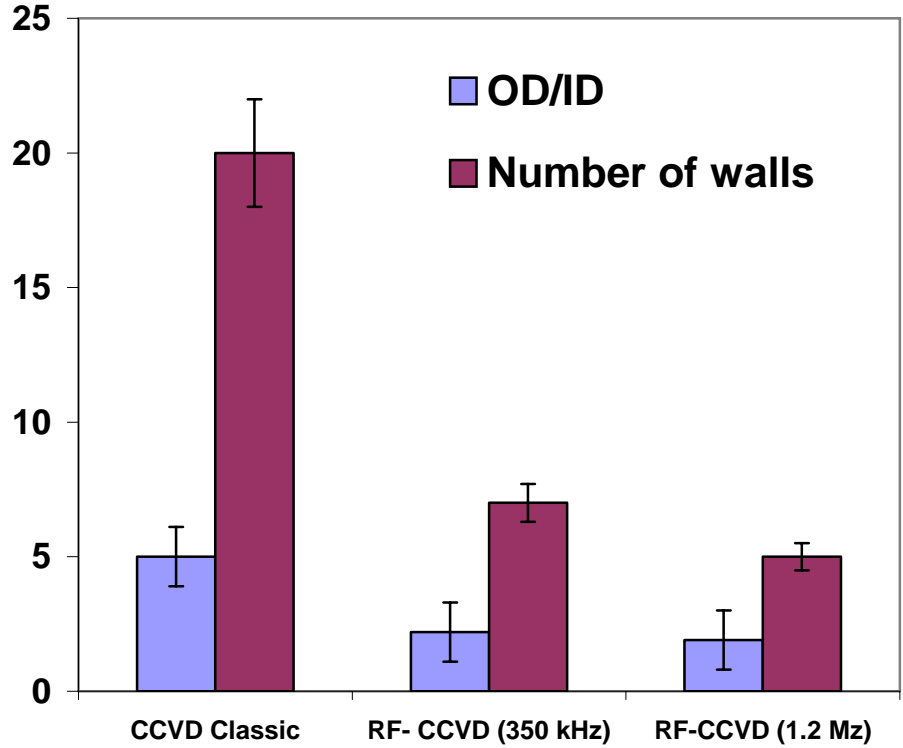

Figure 4. The variation of the ratios of the outside to the inside diameter (OD/ID) and the number of carbon walls for the multiwall carbon nanotubes grown in a regular CCVD oven, and in a RF oven at two different frequencies (350 KH and 1.2 MH). It can be observed the fact that as the nanotubes grown by RF excitation develop fewer number of walls and they show smaller OD/ID values.

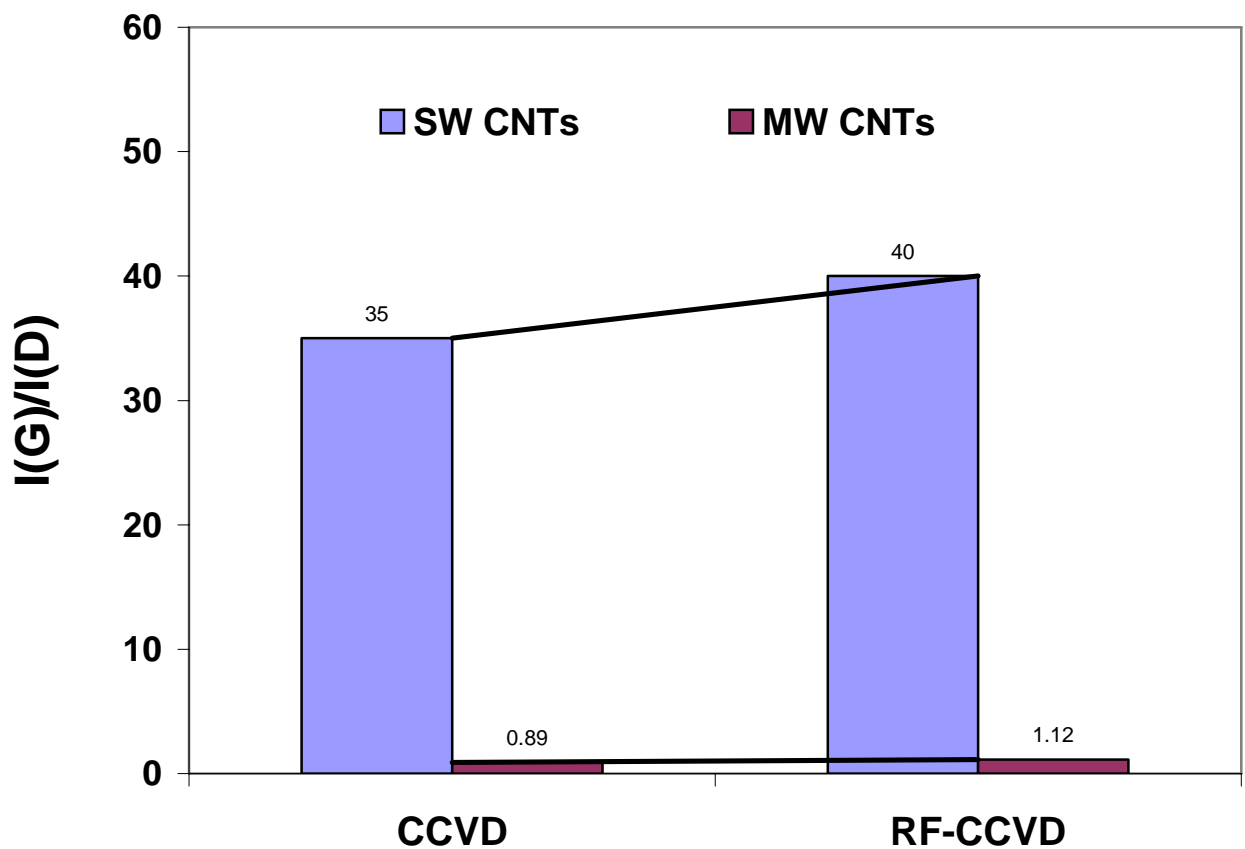

Figure 5. Raman Spectroscopy indicates that the nanotubes (both single and multi walled) show higher crystallinity when grown by RF-CCVD, indicated by higher ratios of the intensities of the G and D bands (I(G)/I(D)).

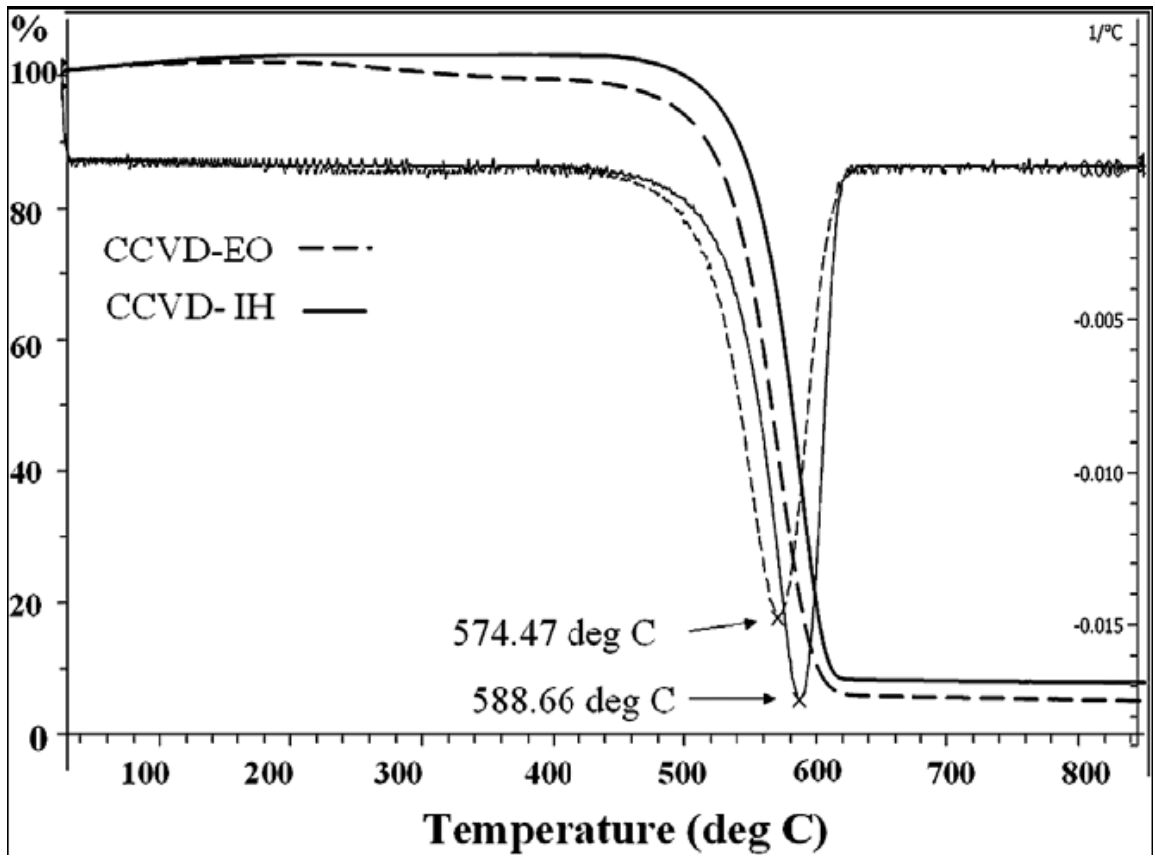

Fig. 6. DTA curves for the multiwall carbon nanotubes synthesized on the Fe:Co:CaCO3 catalyst by CCVD in external oven and by RF excitation. It can be observed that the nanotubes produced by RF excitation show more crystalline properties since they decompose thermally at higher temperatures.

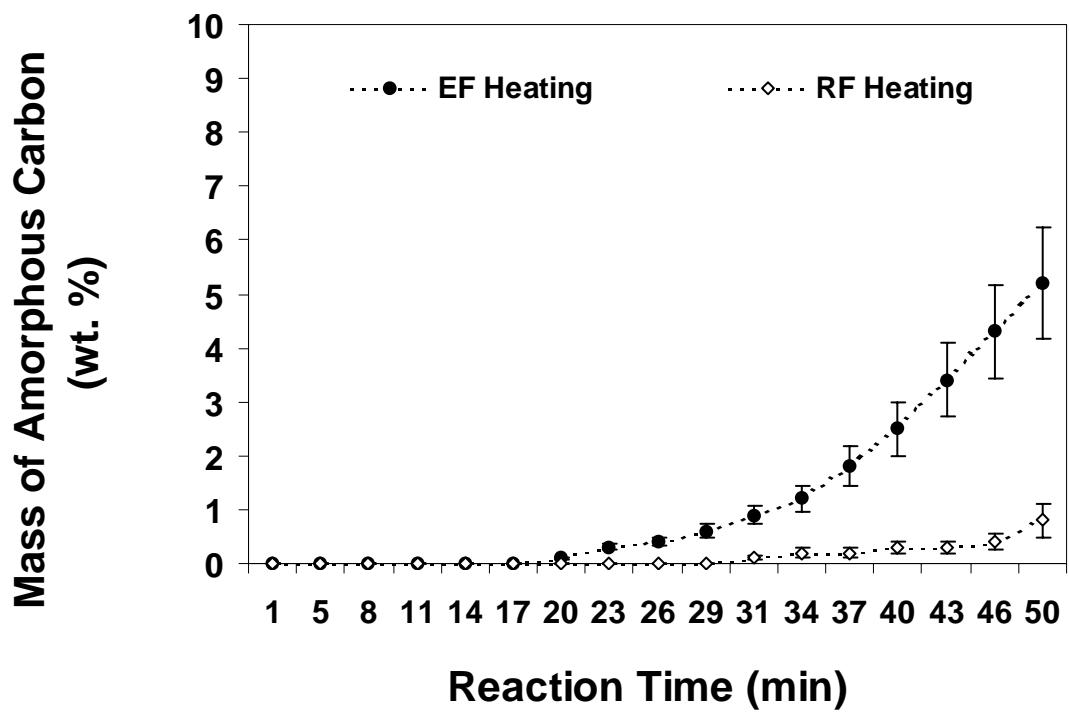

Figure 7. The amount of amorphous carbon generated as a function of reaction time for the two different (EF and RF) heating methods These experimental values indicate that RF is responsible for a more significant arrangement of the carbon atoms into the more crystalline nanotubes and not in the amorphous form.